\documentclass{emulateapj}
\shorttitle{Prompt GeV-TeV Emission of GRBs} 
\shortauthors{Asano \& Inoue}

\begin{document}

\title{
Prompt GeV-TeV Emission of Gamma-Ray Bursts Due to
High-Energy Protons, Muons and Electron-Positron Pairs
}
\author{\scshape Katsuaki Asano and Susumu Inoue}
\affil{Division of Theoretical Astronomy, National Astronomical Observatory of Japan,
2-21-1 Osawa, Mitaka, Tokyo 181-8588, Japan}
\email{asano@th.nao.ac.jp, inoue@th.nao.ac.jp}


\begin{abstract}
In the framework of the internal shock scenario,
we model the broadband prompt emission of gamma-ray bursts (GRBs) with emphasis on the GeV-TeV bands,
utilizing Monte Carlo simulations that include various processes
associated with electrons and protons accelerated to high energies.
While inverse Compton emission from primary electrons is often dominant,
different proton-induced mechanisms can also give rise to distinct high-energy components,
such as synchrotron emission from protons, muons
or secondary electrons/positrons injected via photomeson interactions.
In some cases, they give rise to double spectral breaks 
that can serve as unique signatures of ultra-high-energy protons.
We discuss the conditions favorable for such emission,
and how they are related to the production of ultra-high-energy cosmic rays and neutrinos in internal shocks. 
Ongoing and upcoming observations by {\it GLAST}, atmospheric Cerenkov telescopes and other facilities
will test these expectations and provide important information on the physical conditions in GRB outflows.
\end{abstract}


\section{Introduction}
\label{sec:intro}
The prompt emission of gamma-ray bursts (GRBs) is characterized by
rapid temporal variability and nonthermal spectra extending to high energies,
implying an origin in ultrarelativistic outflows with bulk Lorentz factors $\Gamma \gtrsim100$
\citep[see, e.g., reviews by][]{pir05,mes06}.
In the widely discussed internal shock scenario,
collisions among inhomogeneities within the flow lead to formation of shocks
that convert bulk kinetic energy into Fermi-accelerated, power-law distributions of relativistic electrons,
which then emit synchrotron photons to be observed as the MeV-range gamma-rays \citep{ree94}.
However, a number of challenges for the internal shock model have been pointed out
concerning the radiative efficiency, low energy spectral slope,
various kinds of luminosity correlations, etc.,
and very different alternative models have been proposed
\citep[and references therein]{pir05,mes06,fox06}.
In order to unravel the true nature of the prompt emission
as well as to constrain important physical quantities such as $\Gamma$ and magnetic fields in the outflow,
more broadband observations including the GeV-TeV bands are warranted. 

The physical conditions inferred for internal shocks indicate that
protons may be Fermi-accelerated to energies $\sim 10^{20}$ eV,
making GRBs potential sources of the observed ultra-high-energy cosmic rays \citep[UHECRs;][]{wax95,vie95}.
To test the GRB origin of UHECRs and distinguish it from other possibilities \citep{tor04,ino07},
it is essential to search for characteristic, UHE proton-induced signatures of secondary neutral radiation
that can be observed in coincidence with GRBs.
Besides production of high-energy neutrinos \citep[and references therein]{wax97,mes06b},
efficient proton acceleration may induce
distinctive emission components in the GeV-TeV bands
\citep[and references therein]{zha04,pir05,mes06,der06}.

So far, observational information on GRB GeV-TeV emission has been quite limited.
The {\it EGRET} instrument onboard {\it CGRO} was able to detect GeV emission
from just a handful of the brightest bursts \citep{hur94,din01,gon03}. 
No strong evidence of emission in the TeV region has been found to date
\citep[e.g.][]{con97,atk05,alb07,hor07},
but this could be largely due to the generally high redshifts of GRBs
and the consequent attenuation by pair production with extragalactic background radiation \citep[e.g.][]{man96}.

However, significant advances are expected soon with the launch of {\it GLAST}\footnote{http://glast.gsfc.nasa.gov/},
with greatly improved sensitivity and wider field of view at GeV energies.
TeV emission from bursts at sufficiently low redshift
may eventually be discovered through ongoing observations with current Cerenkov telescopes such as
H.E.S.S.\footnote{http://www.mpi-hd.mpg.de/hfm/HESS/HESS.html},
VERITAS\footnote{http://veritas.sao.arizona.edu/},
CANGAROO III\footnote{http://icrhp9.icrr.u-tokyo.ac.jp/},
and especially MAGIC\footnote{http://magic.mppmu.mpg.de/}
with its 50 GeV threshold and fast slewing capabilities,
as well as all-sky detectors such as MILAGRO\footnote{http://www.lanl.gov/milagro/}.

In anticipation of the observational progress,
this paper discusses detailed theoretical modeling of GRB prompt emission
in the context of the internal shock scenario, focusing on the GeV-TeV bands.
Monte Carlo techniques are employed 
to account for cascade processes involving
photon-photon ($\gamma \gamma$) pair production and Klein-Nishina regime Compton scattering,
as well as proton-induced processes such as
photomeson interactions and secondary pion, muon, electron and positron injection.
Although various aspects of high energy emission from internal shocks
have been covered in previous studies \citep[e.g.][]{pap96,pil98,gue03,pee04b,raz04,bar06},
few have discussed hadronic cascade processes in such detail.

In \S \ref{sec:model}, our model assumptions, methods and choice of parameters are explained.
\S \ref{sec:cutoffic} summarizes some general aspects of the high-energy cutoff and inverse Compton emission.
The effects induced by high-energy protons are highlighted in \S \ref{sec:proton},
and the relation between GeV-TeV emission and UHECR and neutrino production is discussed in \S \ref{sec:uhe}.
We briefly touch on the observational implications in \S \ref{sec:obs},
and conclude in \S \ref{sec:conc}.

\section{Model Description}
\label{sec:model}

\subsection{Model Assumptions and Numerical Methods}
\label{sec:assum}

In the internal shock picture, each pulse observed in the MeV light curve of GRBs
is interpreted as emission from shocks formed in collisions
between material travelling at different velocities \citep{kob97,dai98}.
Here we do not deal with the dynamics of the shocks
and instead concentrate on the emission properties.
The emitting region for a pulse is considered to be a homogeneous shell
expanding with $\Gamma$ at radii $R$ from the central engine.
We adopt $l=R/\Gamma$ for the comoving width of the shell,
so that the pulse timescale in the observer frame is $\Delta t=R/\Gamma^2 c$
\citep{sar97},\citep[see however][]{asa02}.
Note that our spherically symmetric formulation is equally valid for a collimated outflow
so long as the collimation angle $\gg 1/\Gamma$.

Detailed modeling of the GRB spectra including the rapid, irregular time variability
would entail considerable complexity.
In this work, we choose not to consider the time variability in earnest and assume steady state conditions,
at least during the pulse timescale $\Delta t$. 
For bursts composed of multiple pulses,
we also assume for simplicity that all pulses within a burst are similar,
i.e. they are emitted from $N$ shells with identical physical conditions.
Our results are therefore to be interpreted as the time-averaged spectra for each burst.

We employ the Monte Carlo numerical code of \citet{asa05} and \citet{asa06},
newly supplemented with $\gamma \gamma$ pair production and synchrotron self-absorption. 
All photons and particles (electrons, positrons, protons, pions, muons)
are distributed isotropically in the shell frame and treated in the one-zone approximation.
Being mutually affected through processes such as
photomeson interactions and inverse Compton (IC) scattering, 
the energy distributions of photons and particles are simulated iteratively
until they converge to a self-consistent steady state,
which is assumed to be realized within the pulse timescale.

The energy density of accelerated electrons in the shell $U_{\rm e}$
is a parameter that can be directly related to observables (\S \ref{sec:param}).
The magnetic field strength $B$ is parameterized by $f_B$
so that its energy density $U_B \equiv B^2/8 \pi=f_B U_{\rm e}$.
Electrons are injected with a power-law energy distribution
$N(\gamma_{\rm e}) \propto\gamma_{\rm e}^{-p_{\rm e}}$ 
in the range $\gamma_{\rm e,min} \le \gamma_{\rm e} \le \gamma_{\rm e,max}$,
where $\gamma_{\rm e}$ is the electron Lorentz factor in the shell frame.
The minimum Lorentz factor $\gamma_{\rm e,min}$ is often evaluated in the literature
by giving $U_{\rm e}$ together with the total number density $n_{\rm e}$ of electrons in the shell,
which can be related to the dissipated kinetic energy \citep[e.g.][]{kob97}.
Instead of considering $n_{\rm e}$,
here we take $\gamma_{\rm e,min}$ to be an additional parameter,
the value of which can be inferred from the observed spectral peak energy (\S \ref{sec:param}).
The maximum Lorentz factor $\gamma_{\rm e,max}$ is where
synchrotron and IC losses limit Fermi acceleration.
However, its value is not very crucial here,
since our choice of $p_{\rm e}$ below (\S \ref{sec:param}) implies
that other factors are more important in shaping the high energy spectra.

Accelerated protons with energy density $U_{\rm p}$
are also injected with a power-law energy distribution 
$\propto \gamma_{\rm p}^{-p_{\rm p}} (\gamma_{\rm p,min} \le \gamma_{\rm p} \le \gamma_{\rm p,max})$
in the shell frame. 
The maximum proton Lorentz factor $\gamma_{\rm p, max}$ is determined by
equating $t_{\rm acc}= \gamma_{\rm p} m_p c^2/e B c$,
the Fermi acceleration timescale in relativistic shocks \citep[e.g.][]{wax95},
to $\min[t_{\rm exp}, t_{\rm loss}]$, where
$t_{\rm exp}=R/\Gamma c$ is the comoving expansion timescale and
$t_{\rm loss}$ is the energy loss timescale due to synchrotron, IC, and photomeson cooling,
as described in \citet{asa05}.
The minimum proton Lorentz factor $\gamma_{\rm p, min}$ is expected to be of order unity
in internal shocks with typically mildly relativistic velocities;
here we take $\gamma_{\rm p, min}=10$, although the exact value is irrelevant for the resulting spectra.

As in \citet{asa06},
we utilize experimental results for the cross sections of the reactions
$p \gamma \to n \pi^+$, $p \pi^0$, $n \pi^+ \pi^0$, and $p \pi^+ \pi^-$ for $\varepsilon' \leq 2$ GeV,
where $\varepsilon'$ is the photon energy in the proton rest frame \citep{sch03}.
The process $p \gamma \to p \pi^0 \pi^0$ is neglected due to its small cross section.
For pion production by $n \gamma$ reactions,
we adopt the same cross sections as the respective $p \gamma$ channels.
The inelasticity is approximated by $K=[1-(m_{\rm p}^2-m^2)/s]/2$,
where $s$ is the center-of-momentum energy squared for the $p \gamma$ or $n \gamma$ system,
$m=m_\pi$ and $m=2m_\pi$ for single and double pion production, respectively,
and $m_\pi$ is the pion mass.
Pion production via $pp$ collisions is not considered here
since target photons always greatly outnumber protons.

We account for the decay of pions and muons
and associated electron/positron injection
as well as synchrotron and IC emission from all charged particles
with the methods of \citet{asa05}.
The full Klein-Nishina cross section \citep[e.g.][]{blu70} is employed for IC scattering.
For synchrotron radiation from very high-energy electrons/positrons, quantum effects can become important.
When the classical value for the synchrotron photon energy
$\varepsilon_{\rm syn}=\gamma_{\rm e}^2 \hbar e B/m_{\rm e} c$
is larger than 10\% of the particle energy $\gamma_{\rm e} m_{\rm e} c^2$,
we use approximate emissivity formulae following \citet{erb66}.
The details of this treatment do not affect the results significantly,
as such synchrotron photons promptly create further pairs 
and the initial information is lost in the cascade process.
For the same reason, we also do not distinguish between
the cascade contributions from pions and muons.

Newly implemented here into the Monte Carlo code with the appropriate cross sections
are $\gamma \gamma$ pair production and synchrotron self-absorption by electrons/positrons.
The cross section for $\gamma \gamma$ pair production is
$\sigma_\pm=\sigma_{\rm T} g(y)$,
where $\sigma_{\rm T}$ is the Thomson cross section,
\begin{eqnarray}
g(y) \equiv \frac{3}{16} (1-y^2) \left[
(3-y^4) \ln{\frac{1+y}{1-y}}-2 y (2-y^2) \right] ,
\end{eqnarray}
$y$ is given by
$y^2=1-(2 m_{\rm e}^2 c^4)/[\varepsilon_1 \varepsilon_2 (1-\cos{\theta})]$,
$\varepsilon_1$ and $\varepsilon_2$ are the energies of the two photons
and $\theta$ is their incident angle \citep{ber82}.
For synchrotron absorption of an isotropic photon field by electrons/positrons,
the differential cross sections for true absorption and stimulated emission are respectively
\begin{eqnarray}
\frac{d \sigma_{\rm a}}{d \Omega} (\gamma_{\rm e},\varepsilon_0)&=&
\frac{c^2 h^3 \gamma'_{\rm e} u'_{\rm e}}
{8 \pi \varepsilon_0^3 \gamma_{\rm e} u_{\rm e}}
P(\gamma'_{\rm e},\varepsilon_0), \\
\frac{d \sigma_{\rm s}}{d \Omega} (\gamma_{\rm e},\varepsilon_0)&=&
\frac{c^2 h^3}
{8 \pi \varepsilon_0^3}
P(\gamma_{\rm e},\varepsilon_0),
\end{eqnarray}
where $\varepsilon_0$ is the photon energy, 
$\gamma'_{\rm e}=\gamma_{\rm e}+\varepsilon_0/m_{\rm e} c^2$,
$u_{\rm e}=(\gamma^2_{\rm e}-1)^{1/2}$, $u'_{\rm e}=({\gamma'}^2_{\rm e}-1)^{1/2}$,
and $P(\gamma_{\rm e},\varepsilon_0)$ is the synchrotron power per unit photon energy \citep{ghi91}.
An accurate treatment of synchrotron self-absorption is necessary
to determine the correct photon spectrum at very low energies,
which in turn is essential for properly evaluating the photomeson interaction rate for UHE protons.

We do not include pair annihilation, which can lead to a prominent spectral component
for sufficiently high compactness parameters,
but only in a narrow energy range around $\Gamma m_{\rm e} c^2$ \citep{pee04b}.

\subsection{Constraints on Parameters}
\label{sec:param}

The full set of our model parameters consists of
$\Gamma$, $R$, $N$, $U_{\rm e}$, $f_B$, $\gamma_{\rm e,min}$, $p_{\rm e}$,
$U_{\rm p}$ and $p_{\rm p}$. 
For $\Gamma$ and $R$, we consider the ranges 
$\Gamma=30$-$1000$ and $R=10^{13}$-$10^{16}$ cm, 
as generally discussed for internal shock models \citep[e.g.][]{mes00}.
We assume a range of $f_B=0.1$-$30$ for the magnetic field (see below).
In order to keep the scope of the current study tractable,
some combinations of the remaining parameters are constrained
so as to reproduce typically observed properties of the MeV, primary synchrotron component.

For given values of $B$ and $\Gamma$,
$\gamma_{\rm e,min}$ is chosen so that the corresponding synchrotron photon energy in the observer frame
$\varepsilon_{\rm pk} = \Gamma \gamma_{\rm e,min}^2 \hbar e  /m_{\rm e} c$
is always 300 keV (for GRB redshift $z=0.1$, see below).
The electron injection index is fixed to $p_{\rm e}=3$,
implying that in the fast-cooling conditions of internal shocks,
the photon index immediately above $\varepsilon_{\rm pk}$
is $\beta=-(p_{\rm e}+2)/2=-2.5$, the mean value measured by BATSE \citep{pre00}.

During the pulse timescale $\Delta t$, 
the fast-cooling electrons reach steady state where 
$U_{\rm e} \simeq U_{\gamma, e}$, the energy density of
photons emitted by electrons in the shell rest frame.
The isotropic-equivalent energy of photons from a single pulse is thus
$E_{\rm sh,e} = 4 \pi \Gamma^2 U_{\rm \gamma, e} R^2 c \Delta t \simeq 4 \pi R^3 U_{\rm e}$.
In all cases studied below, the emitted luminosity is dominated by MeV synchrotron photons,
so for given $R$, $U_{\rm e}$ can be related to the observable MeV pulse energy $E_{\rm sh}$.
Hereafter $U_{\rm e}$ is replaced by $E_{\rm sh}$ as a parameter in the range $10^{50}$-$10^{52}$ erg.
Under our assumption of $N$ identical pulses constituting a burst (\S \ref{sec:assum}),
the time-integrated, isotropic-equivalent photon energy for a burst is $E_{\rm tot} = N E_{\rm sh}$,
which we fix to a typical value of $10^{53}$ erg.

Although the proton component cannot be strongly constrained from  
existing observations,
we assume $U_{\rm p} = U_{\rm e}$ and $p_{\rm p}=2$,  
which are necessary conditions
for GRBs to be energetically viable as UHECR sources \citep 
{wax95,vie95}.
(However, recent observations may suggest larger values of $U_ 
{\rm p}$, \S \ref{sec:uhe}.)
The proton spectral index $p_{\rm p}$ is expected to be similar  
to $p_{\rm e}$
at low energies where the particle gyroradii overlap,
but this may not necessarily be the case at ultra-high-energies that  
are important for photomeson interactions.
In particular, if the nonlinear back-reaction of CR pressure on the  
shock structure is significant,
a concave spectral shape may result that is much harder at high  
energies compared to low energies \citep{mal01},
even though the details are uncertain for relativistic shocks  
\citep{bar91}.

After specifying the observables $\varepsilon_{pk}$, $\beta$ and $E_{\rm tot}$ to typical values
and making plausible assumptions for the protons,
the remaining variable parameters are $\Gamma$, $R$, $f_B$ and $E_{\rm sh}$.
Utilizing the observable pulse timescale $\Delta t = R/\Gamma^2 c$ instead of $R$,
we choose to characterize our results with the set of $\Delta t$, $E_{\rm sh}$, $\Gamma$ and $f_B$.
Note that a relation can also be made to the pulse luminosity $L = E_{\rm sh}/\Delta t$.

All spectra below are shown in terms of the observed fluence versus  
photon energy,
assuming a GRB redshift of $z=0.1$.
We do not include spectral attenuation by pair production with the  
extragalactic infrared background,
which may be justified at $z \la 0.1$ and $ 
\varepsilon \la 3$ TeV \citep{aha06},
but should be more important for higher redshifts and photon energies.
The potential effects of intergalactic cascade emission \citep 
[e.g.][]{pla95,raz04,wan04,cas07,mur07}
are also neglected.

We caution that actual GRBs are observed with considerable dispersions
in $\varepsilon_{\rm pk}$, $\beta$ and $E_{\rm tot}$,
notwithstanding a good correlation between $\varepsilon_{\rm pk}$ and $E_{\rm tot}$ \citep{ama06}.
Pulses within each burst can also exhibit a variety of properties.
Such aspects need to be accounted in future, more comprehensive studies.

Note that cases of $f_B=U_B/U_e \gg 1$
can be compatible with internal shocks in a kinetic energy-dominated outflow,
as long as the fraction of protons and electrons injected into the acceleration process is sufficiently small,
and most of the outflow energy remains in the form of cold or thermal protons.
Indeed, the typical radiative efficiency expected from electrons accelerated in internal shocks
is only a few percent \citep[e.g][]{dai98} \citep[see however][]{zha07},
so that $f_B$ as large as 30 may still be consistent with this picture.
Even in magnetically-dominated flows, shocks can occur under certain conditions \citep{zha05}.


\section{High-Energy Cutoff and Inverse Compton Emission}
\label{sec:cutoffic}

Before proton-induced effects are addressed in detail in \S \ref{sec:proton},
we discuss some generic aspects of the high-energy spectral cutoff
that are independent of the emission mechanism,
together with the properties of GeV-TeV spectra in the typical case
where inverse Compton emission from electrons dominate.

\subsection{High-Energy Cutoff}
\label{sec:cutoff}

\begin{figure}[h]
\vspace{1.5cm}
\centering
\epsscale{0.9}
\plotone{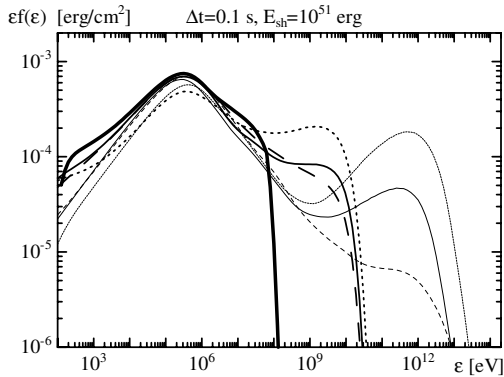}
\caption{
Spectra for $\Delta t=0.1$ s, $E_{\rm sh}=10^{51}$ erg 
and varying $\Gamma$ and $f_B$.
The thickest curve is for $\Gamma=100$ and $f_B=1.0$.
The medium-thick and thin curves are for $\Gamma=$300 and 1000, respectively,
while dotted, solid and dashed each correspond to $f_B=$0.1, 1, and 30.
\label{fig:gammab}}
\end{figure}

In Fig. \ref{fig:gammab}, we show exemplary spectra
for the case of $\Delta t = 0.1$ s, $E_{\rm sh}=10^{51}$ erg,
and different values of $\Gamma$ and $f_B$.
Above the synchrotron peak at $\varepsilon_{\rm pk}=$ 300 keV,
there are varying levels of a second high-energy component,
here all due to inverse Compton emission.
Clear spectral cut-offs can be seen at the highest energies,
above where pair production with low energy photons
within the emission region strongly attenuates the spectrum.

The high-energy cutoff energy $\varepsilon_{\rm cut}$
should provide an effective probe of the bulk Lorentz factor $\Gamma$,
as has been discussed previously \citep[e.g.][]{bar97,lit01}.
For the case of a pure power-law spectrum,
\citet{asa03b} have obtained an analytical expression,
$\varepsilon_{\rm cut} \propto \Gamma^{14/3} E_{\rm sh}^{-2/3} \Delta t^{4/3}$ for  $p_{\rm e}=3$
(or $\varepsilon_{\rm cut} \propto \Gamma^{26/5} E_{\rm sh}^{-4/5} \Delta t^{8/5}$ for $p_{\rm e}=5/2$).
Our Monte Carlo simulation results reveal values of $\varepsilon_{\rm cut}$
that are too scattered to be fit well by one simple, analogous formula.
Nevertheless, it can be approximated roughly by
\begin{eqnarray}
\varepsilon_{\rm cut} \simeq 10^9 \left( \frac{\Gamma}{100} \right)^{4}
\left( \frac{E_{\rm sh}}{10^{51} {\rm erg}} \right)^{-0.5}
\left( \frac{\Delta t}{1 {\rm s}} \right)^{1.3}
{\rm eV},
\end{eqnarray}
with a typically strong $\Gamma$-dependence.
Thus, together with the observables $\Delta t$ and $E_{\rm sh}$ (or $L$),
measurements of $\varepsilon_{\rm cut}$ should provide tight constraints on $\Gamma$.
These inferences are mostly independent of the emission process that shape the GeV-TeV spectra,
whether IC or not.

\subsection{Inverse Compton Emission}
\label{sec:IC}

Detecting the IC component should be crucial for probing the magnetic field strength,
especially for larger values of $\Gamma$.
For $\Gamma=$300-1000 in Fig. \ref{fig:gammab},
the strong dependence of the IC fluence on $f_B$ is apparent.
For lower $f_B$ and consequently higher IC fluence,
the synchrotron fluence is somewhat suppressed due to the greater importance of IC cooling.
In contrast, for $\Gamma=100$,
the spectra for different $f_B=$0.1-30 are almost indistinguishable,
as $\varepsilon_{\rm cut}$ occurs at too low energies for the IC component to be clearly discerned.
Here the dominant high-energy component is simply 
the extension of the primary synchrotron emission up to $\varepsilon_{\rm cut} \sim 0.1$ GeV.

\begin{figure}[h]
\vspace{1.5cm}
\centering
\epsscale{0.9}
\plotone{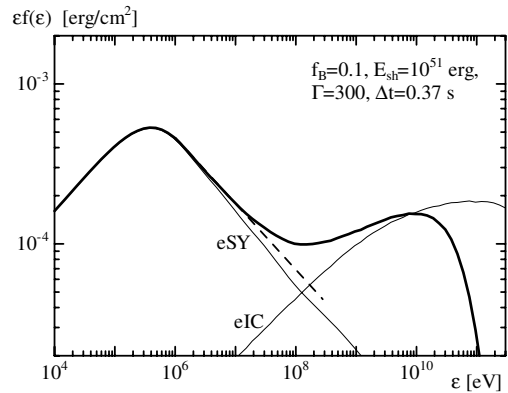}
\caption{
Spectrum for a case with inverse Compton peak (thick solid),
for $\Delta t=0.37$ s, $E_{\rm sh} = 10^{51}$ erg, 
$\Gamma=300$ and $f_B=0.1$.
Thin curves denote separately electron synchrotron (labeled eSY)
and inverse Compton (labeled eIC) components, without $\gamma \gamma$ absorption effects. 
The dashed line is a power-law extrapolation of the MeV-range spectrum.
\label{fig:ic}}
\end{figure}

Fig. \ref{fig:ic} displays more details of the spectra for a case
where IC emission makes a distinct second peak at $\varepsilon \sim 10$ GeV.
The IC to synchrotron peak fluence ratio is less than the simple Thomson limit expectation
$\propto U_{\rm e}/U_B=f_B^{-1}$ because of the Klein-Nishina effect and $\gamma \gamma$ absorption.
Note that ``electrons'' here include both primary electrons, i.e. those directly accelerated at the shocks,
as well as additional pairs that are injected by $\gamma \gamma$ interactions at higher energies. 

These results on IC emission from internal shocks
are broadly consistent with previous, more approximate studies \citep[e.g.][]{pap96,pil98,gue03}.
In \S \ref{sec:icpeak},
we show some quantitative relations for
the spectral peak energies and fluence ratios in our model,
which might offer a useful consistency check of the internal shock scenario,
at least within the parameter space studied here.

\section{Proton-Induced High-Energy Emission}
\label{sec:proton}

\begin{figure}[h]
\vspace{1.5cm}
\centering
\epsscale{0.9}
\plotone{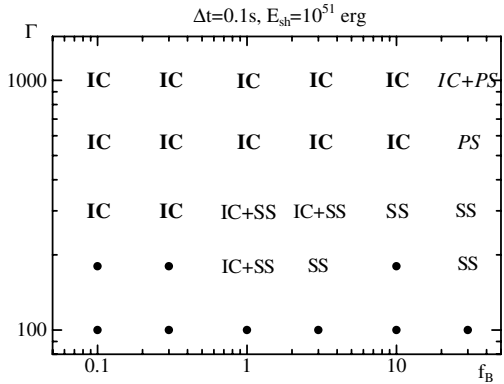}
\caption{
Summary of visible high-energy spectral components in the $f_B$-$\Gamma$ plane,
denoted by IC (inverse Compton), PS (proton synchrotron) and SS (secondary pair synchrotron),
for $\Delta t = 0.1$ s and $E_{\rm sh} = 10^{51}$ erg (spiky pulse case). 
Black dots imply no distinct high-energy feature.
\label{fig:spiky}}
\end{figure}

\begin{figure}[h]
\vspace{1.5cm}
\centering
\epsscale{0.9}
\plotone{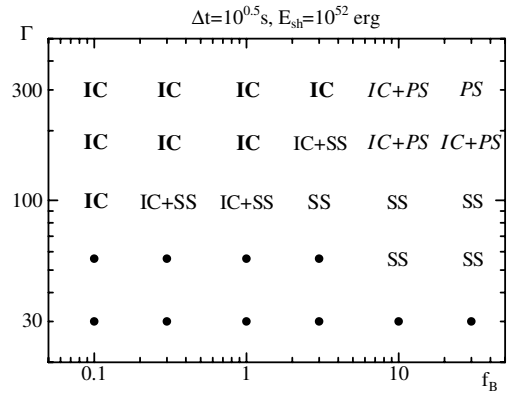}
\caption{
Same as Fig.\ref{fig:spiky},
but for $\Delta t = 10^{0.5}$ s and $E_{\rm sh} = 10^{52}$ erg (broad-pulse case). 
\label{fig:broad}}
\end{figure}

For the range of parameters covered in this study (\S \ref{sec:param}),
IC emission often turns out to be the dominant high-energy emission mechanism
(see Figs. \ref{fig:spiky}, \ref{fig:broad}).
Nevertheless, we find that within plausible parameter regimes,
distinctive spectral features can emerge at GeV-TeV energies
due to characteristic processes induced by UHE protons.

\begin{figure}[h]
\vspace{1.5cm}
\centering
\epsscale{0.9}
\plotone{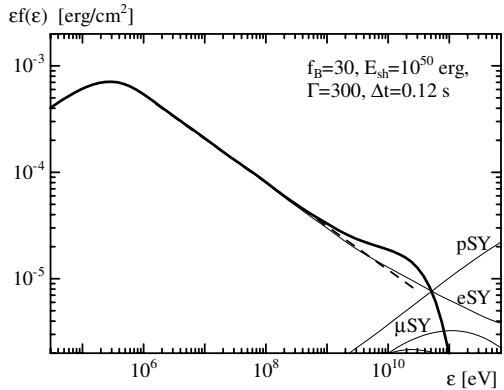}
\caption{
Spectrum for a case with proton synchrotron bump (thick solid),
for $\Delta t = 0.12$ s, $E_{\rm sh} = 10^{50}$ erg, 
$\Gamma=300$ and $f_B=30$.
Thin curves denote separately
electron synchrotron (eSY), proton synchrotron (pSY) and muon synchroton ($\mu$SY) components,
without $\gamma \gamma$ absorption effects. 
The dashed line is a power-law extrapolation of the MeV-range spectrum.
\label{fig:psyn}}
\end{figure}

One potential radiative signature of UHE protons in GRBs
is their synchrotron emission, first proposed by \citet{vie97}.
Fig. \ref{fig:psyn} is a case with relatively high magnetic fields ($f_B=30$),
in which proton synchrotron emission makes a marked contribution
to the spectrum at $\varepsilon \sim$ 1-100 GeV.
Although the spectral bump here is not as prominent as some examples of IC peaks (\S \ref{sec:IC}),
it forms a clear excess above a simple extrapolation of the MeV-band spectrum
that may be detectable by GLAST, MAGIC and other facilities.

\begin{figure}[h]
\vspace{1.5cm}
\centering
\epsscale{0.9}
\plotone{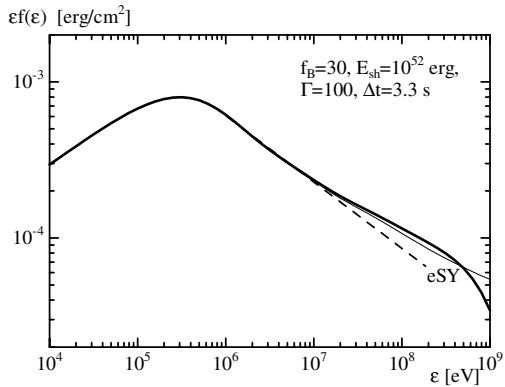}
\caption{
Spectrum for a case with secondary pair synchrotron bump (thick solid),
for $\Delta t = 3.3$ s, $E_{\rm sh} = 10^{52}$ erg, 
$\Gamma=100$ and $f_B=30$.
Thin curve denotes the synchrotron component (eSY) without $\gamma \gamma$ absorption effects. 
The dashed line is a power-law extrapolation of the MeV-range spectrum.
\label{fig:ssyn}}
\end{figure}

Another example with high magnetic fields ($f_B=30$) is shown in Fig. \ref{fig:ssyn}.
Here, a notable hardening of the spectrum can be seen at  $\varepsilon \sim $0.01-1 GeV,
caused by synchrotron emission from secondary pairs injected by photomeson interactions.
Since primary electrons 
alone cannot give rise to such distinct features,
this is a unique effect triggered by UHE protons,
which has not been discussed before for GRB prompt emission.
Note that in order to correctly evaluate the density of low-energy target photons for the $p \gamma$ process,
it is imperative to include self-absorption effects in the electron synchrotron spectrum (\S \ref{sec:assum}).

\begin{figure}[h]
\vspace{1.5cm}
\centering
\epsscale{0.9}
\plotone{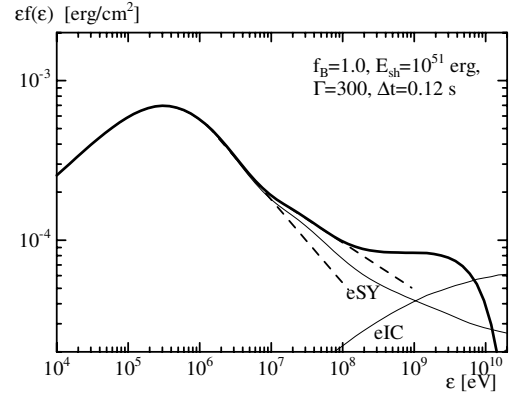}
\caption{
Spectrum for a case with double breaks due to secondary pair synchrotron and IC emission (thick solid),
for $\Delta t = 0.12$ s, $E_{\rm sh} = 10^{51}$ erg, 
$\Gamma=300$ and $f_B=1$.
Thin curves denote separately electron synchrotron (eSY) and IC (eIC) components,
without $\gamma \gamma$ absorption effects. 
The dashed lines are power-law extrapolations of the spectra in the ranges 1-10 MeV and 10-100 MeV.
\label{fig:double}}
\end{figure}

Although such features should serve as valuable indicators of UHE protons in GRBs,
it may not be easy from spectral measurements alone 
to distinguish them from some cases of IC emission.
However, under certain conditions,
more than one emission mechanism can become simultaneously important and lead to double spectral breaks,
which can only occur in the presence of accelerated protons.
Fig. \ref{fig:double} is an example where the spectrum
hardens above a first break at $\sim 0.01$ GeV from secondary pair synchrotron emission,
and then hardens further above a second break at $\sim 0.1$ GeV from IC emission.
Spectra with such double breaks may offer crucial observational evidence for UHE proton acceleration.

\begin{figure}[h]
\vspace{1.5cm}
\centering
\epsscale{0.9}
\plotone{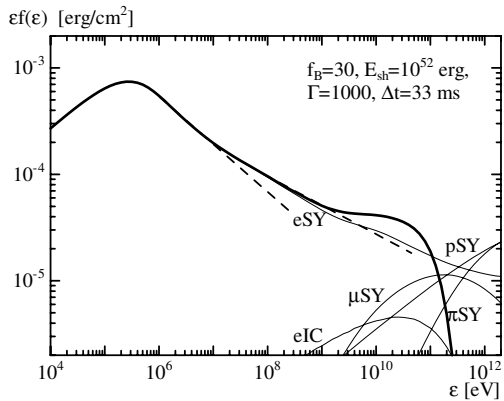}
\caption{
Spectrum for a case with muon synchrotron bump (thick solid),
for $\Delta t = 33$ ms, $E_{\rm sh} = 10^{52}$ erg, 
$\Gamma=1000$ and $f_B=30$.
Thin curves denote separately
electron synchrotron (eSY), IC (eIC), proton synchrotron (pSY),
muon synchroton ($\mu$SY), and pion synchrotron ($\pi$SY) components,
without $\gamma \gamma$ absorption effects. 
The dashed lines are power-law extrapolations of the spectra in the ranges 1-10 MeV and 10-100 MeV.
\label{fig:msyn}}
\end{figure}

Yet a third proton-induced process that can be significant is synchrotron emission from muons
injected by $p \gamma$ interactions, first discussed by \citet{asa03b}.
In Fig. \ref{fig:msyn}, again for high magnetic fields ($f_B=30$),
a muon synchrotron spectral bump is eminent at $\varepsilon \sim$10-100 GeV.
Additionally visible in this case are
secondary pair synchrotron emission at $\varepsilon \sim$  
0.1-1 GeV,
proton synchrotron emission at $\varepsilon \sim$ 100 GeV,
and even a minor contribution from pion synchrotron emission at $ 
\varepsilon \ga$ 100 GeV,
illustrating the spectral variety generated by UHE protons.
While not shown here, there are other instances where muon synchrotron is the sole high-energy component
(see \S \ref{sec:summa}).

It is important to clarify in which physical regimes of GRB internal shocks
these proton-related emission components become clearly visible.
Here we do not attempt to explore the full parameter space,
but choose to map out certain ranges of $\Gamma$ and $f_B$
while focusing on the following two sets of $\Delta t$ and $E_{\rm sh}$:
(1) ``spiky pulse'' case of $\Delta t = 0.1$ s and $E_{\rm sh} = 10^{51}$ erg, 
(similar to Figs. \ref{fig:psyn}, \ref{fig:ssyn} and \ref{fig:double}),
and
(2) ``broad pulse'' case of $\Delta t = 10^{0.5}$ s and $E_{\rm sh} = 10^{52}$ erg. 
Whenever IC, proton synchrotron or secondary pair synchrotron emission
create distinct spectral features over the extrapolated MeV-range spectra,
we correspondingly indicate ``IC'', ``PS'' or ``SS''  in the $f_B$-$\Gamma$ plane
of Figs. \ref{fig:spiky} and \ref{fig:broad}.
When two components occur simultaneously,
they are indicated together with a ``+'' sign,
while black dots signify that no separate high-energy component is discernible.
(Muon synchrotron emission does not become significant in these two cases,
but can be evident for other $\Delta t$ and $E_{\rm sh}$ as in Fig. \ref{fig:msyn}.)

Generally speaking, we see that GeV-TeV emission requires sufficiently large $\Gamma$
regardless of the emission mechanism to avoid $\gamma \gamma$ absorption (\S \ref{sec:cutoff}),
and that larger $f_B$ is more conducive to proton-induced components.
We also see that cases of multiple components can be fairly common.
A more quantitative summary of the model spectra in the current study can be found in \S \ref{sec:summa},
which may provide a guide to searches for proton-induced signals in future observations.

While proton-induced emission can become clearly observable,
in all the cases studied here, it does not lead to conspicuously separate spectral peaks as for the IC emission.
On the other hand, such situations may be possible outside of the parameter restrictions we set in \S \ref{sec:param}.
For example, some recent observations may point to $U_{\rm p} > U_{\rm e}$
if GRBs are the origin of UHECRs (\S \ref{sec:uhe}).
All proton-related components will then be duly increased,
as they simply scale in proportion to $U_{\rm p}$. 

Although not explicitly addressed in this work,
we mention that time variability should also be a crucial diagnostic of emission mechanisms,
as each process has its characteristic timescale and dependence on photon energy.
This will be an important subject for future studies.

\section{Relation to Ultra-High-Energy Cosmic Ray and Neutrino Production}
\label{sec:uhe}

We now discuss how the above results on GeV-TeV emission
are related to the processes of UHECR and neutrino production in GRB internal shocks.
Since the internal shock model entails a wide range of physical conditions by design,
the circumstances most favorable for each process are not necessarily the same.
We concentrate below on some representative cases
without investigating the full model parameter space.

The acceleration of protons to ultra-high energies is a necessary but not sufficient condition
for GRB internal shocks to be significant contributors of UHECRs,
since the particles must also escape efficiently 
without suffering significant energy losses.
Although a detailed description of UHECR escape is beyond the scope of this paper,
following \citet{asa05},
we can impose a relevant constraint that the particles in question,
say, protons with energy $\varepsilon_{\rm p} \geq 10^{19}$ eV,
do not lose more than half of their energy radiatively in a comoving expansion timescale
after their injection in the shell 
(roughly $t_{\rm loss} \ga  t_{\exp}$).
Strictly speaking, this is a minimum requirement,
but if it is satisfied, we may expect that the higher energy particles can eventually escape
as the shell expands and both the photon density and magnetic field drop rapidly (note $B \propto R^{-3/2}$).
We can then infer a lower bound on the shell radius $R$,
or equivalently on $\Gamma$ for given $\Delta t$ and $E_{\rm sh}$.
From our numerical results, this criterion for efficient UHECR production is approximately
$R \ga 10^{14} (E_{\rm sh}/10^{50} {\rm erg})^{0.5} (\Gamma/300)^{-1}$ cm
or $\Gamma \ga 300 (\Delta t/0.1{\rm s})^{-0.3} (E_{\rm sh}/10^{51} {\rm erg})^{0.2}$.
The dependence on the magnetic field parameter $f_B$ is weak
as the losses are mostly due to photomeson interactions rather than synchrotron radiation
(although the latter becomes more important for higher energies $\varepsilon_{\rm p} \ga 10^{20}$ eV).
Note that effective particle escape via neutron conversion is included in this criterion
and only occurs in a narrow parameter range near the lower limit values.

Turning to gamma-ray emission,
we saw that distinct GeV-TeV components mandate high values of $\Gamma$
from $\gamma \gamma$ optical depth constraints,
irrespective of the emission mechanism (\S \ref{sec:cutoff}).
In fact, the above bound on $\Gamma$ for UHECR production
roughly matches the bound from gamma-rays,
at least for the two exemplary cases of $\Delta t$ and $E_{\rm sh}$ in \S \ref{sec:proton} 
(regions outside the black dots in the $f_B$-$\Gamma$ plane of Figs. \ref{fig:spiky} and \ref{fig:broad}).
Therefore, the appearance of even the IC emission may possibly indicate 
that the physical conditions are also appropriate for efficient UHECR acceleration and escape.
Of course, the emergence of proton-induced emission (only for high $f_B$)
will be most valuable as it can directly probe important quantities
such as $U_{\rm p}/U_{\rm e}$ and $\gamma_{\rm p, max}$.

Detection of high-energy neutrinos is often emphasized
as a definitive observational test of the GRB origin of UHECRs \citep{hal02}.
However, the situation most advantangeous for neutrino production is that
UHE protons undergo efficient photomeson interactions in dense radiation fields without escaping.
This favors small values of $R$ or $\Gamma$ that are
contrary to and almost mutually exclusive with the UHECR criterion,
as shown in \citet{asa05} \citep[see also][]{gia05}.
For example, the requirement that the emitted neutrino fluence $>10^{-5}\ {\rm erg\ cm^{-2}}$ in the current model
corresponds well with $\Gamma \la 300 (\Delta t/0.1{\rm s})^{-0.3} (E_{\rm sh}/10^{51} {\rm erg})^{0.2}$,
entirely the opposite of the UHECR bound above.
(See \S \ref{sec:neutrino} for a summary of the neutrino spectra in the current model.)
Taking this constraint at face value,
we can find some overlap with the lowest $\Gamma$ cases with GeV-TeV components
in Figs. \ref{fig:spiky} and \ref{fig:broad}.
Indeed, the pertinent process is found to be secondary pair synchrotron emission,
which is generated together with neutrinos in $p \gamma$ interactions.
Yet there is also a large parameter space with even lower $\Gamma$
that allows copious neutrino emission but very little gamma-ray or UHECR production.
Although neutrino observations will still be indispensable to verify
that UHE proton acceleration actually occurs in GRBs,
the bursts that emit the most neutrinos
may not be the ones that contribute the most UHECRs.
\citep[Such remarks do not apply if UHECR acceleration can occur in external shocks;][]{vie95,wax00,der02}
\citep[see however][regarding external forward shocks.]{gal99,mil06}

Thus we find that the connection between UHECR, neutrino, and gamma-ray production in GRB internal shocks
is very intimate, but not one-to-one and nontrivial \citep[see also][]{der07}.
Further studies are warranted for a more complete understanding,
but this point should be important to bear in mind for the respective observations.

We remark that all of the above discussion is based on the assumption $U_{\rm p} = U_{\rm e}$ (\S \ref{sec:assum}).
However, recent, post-SWIFT observations reveal the GRB redshift distribution
to be skewed to higher $z$ than previously believed \citep{jak06}.
This may suggest that a larger energy budget with $U_{\rm p} > U_{\rm e}$
may be necessary for GRB UHECR scenarios to remain viable,
implying correspondingly higher gamma-ray and neutrino contributions.
\citep[Note that extreme values such as $U_{\rm p} \sim 10^3 U_{\rm e}$ have also been proposed;][]{tot98}.

\section{Observational Implications}
\label{sec:obs}

Here we briefly comment on the implications for existing and future observations.

Some EGRET-detected GRBs exhibited GeV emission coinciding with the prompt emission,
with spectra that are mostly consistent with an extrapolation of the MeV spectra \citep{din01}.
For GRB 940217, there is some evidence of a separate high-energy component during the prompt phase,
and perhaps in the delayed, hour-timescale emission as well \citep{hur94}\citep[see also][]{der05}.
While the latter is likely to be associated with the external shock \citep[e.g.][]{mes94,boe98,zha01,ino03},
the former could possibly be related to some of the emission processes discussed here.
More information is necessary to be conclusive, however.
A markedly distinct component with a hard spectrum above several MeV was seen in GRB 941017 \citep{gon03},
but the fact that it varied on considerably longer timescales compared to the sub-MeV emission 
may favor an external shock origin \citep[e.g.][]{gra03,pee04a,der04,bel05}.
At any rate, much more detailed studies of the GeV prompt emission
should become feasible soon after the launch of GLAST,
which may detect some or all of the emission components discussed here.

Although clear detections have yet to be achieved at TeV energies,
the MAGIC telescope has conducted rapid follow-up observations for selected GRBs,
in some cases overlapping with the prompt emission phase \citep{alb06,alb07}.
The obtained upper limits reach fluence levels of $< 10^{-7} {\rm erg\ cm^{-2}}$ at $\sim$ 0.1 TeV
with integration times of several minutes, so our fiducial $z=0.1$ burst should be readily detectable.
Estimating the amount of intergalactic attenuation with the baseline background model of \citet{kne04},
MAGIC may be able to detect
the proton synchrotron emission of Fig. \ref{fig:psyn} or muon synchrotron emission of Fig. \ref{fig:msyn}
out to $z \la 1$, and the IC emission of Fig. \ref{fig:ic} to somewhat higher $z$,
approaching the typical redshifts of GRBs.
Thus the prospects are very promising for further observations by MAGIC
as well as other Cerenkov telescopes such as H.E.S.S., VERITAS and  
CANGAROO III,
and especially the near-future upgraded facilities MAGIC II and  
H.E.S.S. II with their lower energy thresholds.

Weak evidence of TeV photons coincident with GRBs 
have also been reported by some surface detectors, e.g. MILAGRITO \citep{atk00}.
However, the inferred energy fluxes are much higher than at MeV,
which is difficult to explain in the current model framework
unless extreme parameters are invoked, e.g. $U_{\rm_p} \gg U_{\rm e}$ \citep{tot98}.
More observations are anticipated for such facilities with their wide-field monitoring capabilities,
including air shower arrays like ARGO-YBJ \citep{dig04}
and even the Pierre Auger Observatory \citep{all05}.

\section{Conclusions and Outlook}
\label{sec:conc}

Following the internal shock scenario and focusing on GeV-TeV energies,
we have modelled the broadband spectra of GRB prompt emission 
through detailed Monte Carlo simulations
including a wide variety of physical processes related to high-energy electrons and protons.
Besides electron inverse Compton emission,
it was shown that interesting proton-induced components
such as proton synchrotron, muon synchrotron and secondary pair synchrotron emission can become clearly visible.
Multiple component spectra with double breaks may offer unique evidence of ultra-high-energy proton acceleration.
The conditions favorable for GeV-TeV emission may also imply
efficient UHECR acceleration and escape, but not necessarily strong neutrino emission.

The observational prospects are very promising for GLAST,
Cerenkov telescopes such as MAGIC (II), H.E.S.S. (II), VERITAS and CANGAROO III,
as well as wide-field surface detector facilities.
Such observations should test the internal shock model of the prompt emission,
provide new insights into the physics of GRB outflows and central engines,
and probe the origin of UHECRs.

Note that since we did not explicitly treat the dynamics of internal shock formation,
some aspects of our study may also be valid in more general scenarios,
e.g. models involving magnetic energy dissipation \citep[e.g.][and references therein]{fox06},
if electrons and protons can be accelerated with similar energy distributions.

Our Monte Carlo simulations have also allowed detailed studies of the prompt optical emission,
with interesting new results concerning both electron- and proton-induced spectral components.
This will be the subject of a separate paper (Asano \& Inoue, in preparation).

More detailed and comprehensive investigations of the current problem
should accommodate a wider range of parameters,
including dispersions in $\varepsilon_{\rm pk}$, $\beta$ and $E_{\rm tot}$,
as well as the variety of pulse properties
\citep[e.g.][]{zha02,asa03a}.
Accounting for time variability is also an important goal for the future.
We note that such effects may potentially smooth out
some of the more subtler spectral features discussed here
and hinder their observational discrimination,
except for the case of sufficiently bright brights where time-resolved spectra can be acquired.
Our detailed formalism for calculating complicated hadronic  
interactions and pair cascades
should also be useful for other applications,
such as high-energy emission from the afterglow phase
\citep {boe98,der99,sar01,zha01,pee05}.

As this work was being completed, we became aware of a preprint by \citet{gup07}
that addresses issues similar to this paper, albeit it with a simpler, analytic formulation.


\acknowledgements
We thank F. Aharonian, Z. Bosnjak, E. Parizot, M. Teshima and especially F. Daigne
for very valuable discussions.

\appendix
\section{Inverse Compton Spectra}
\label{sec:icpeak}

\begin{figure}[h]
\centering
\epsscale{0.45}
\plotone{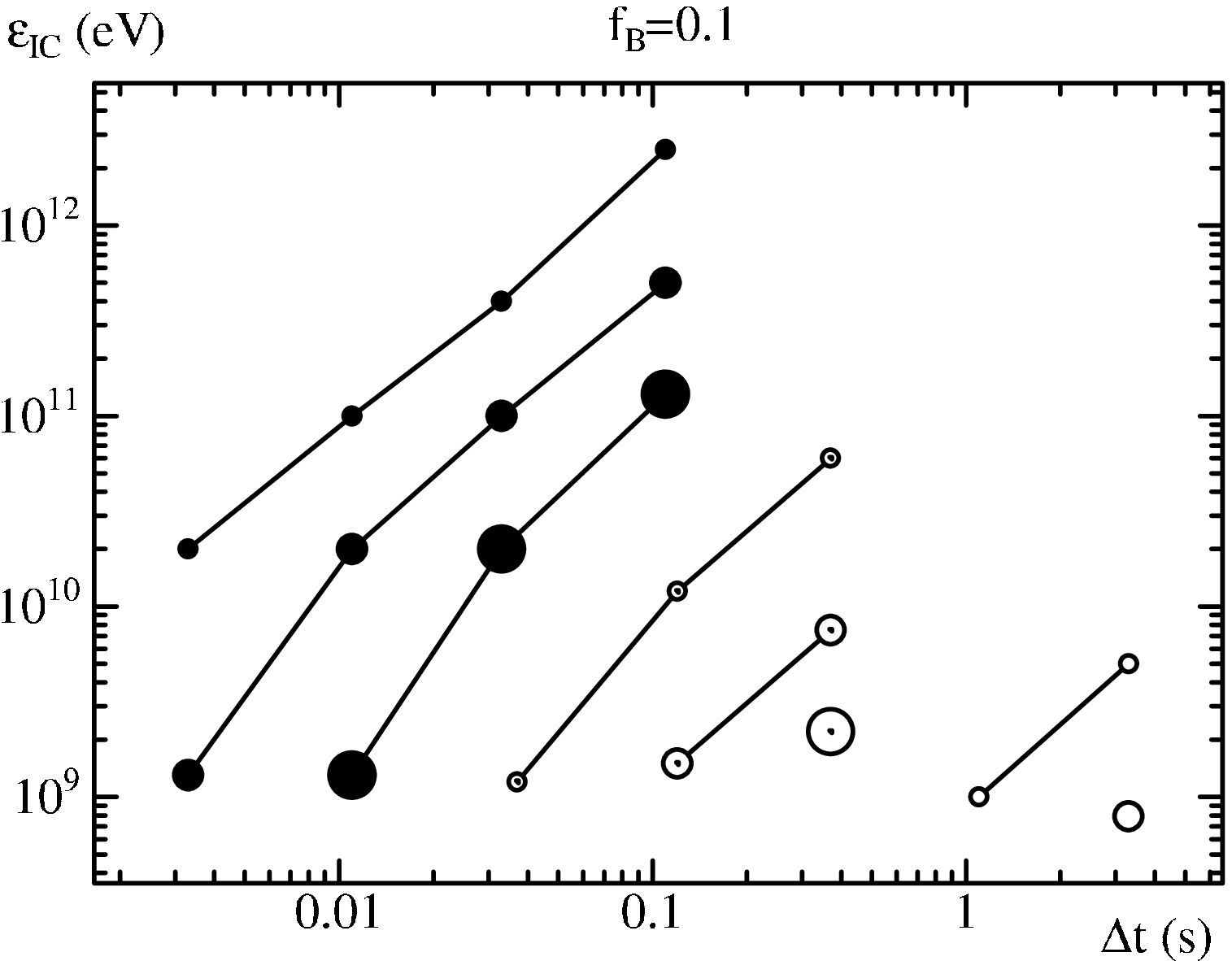}
\plotone{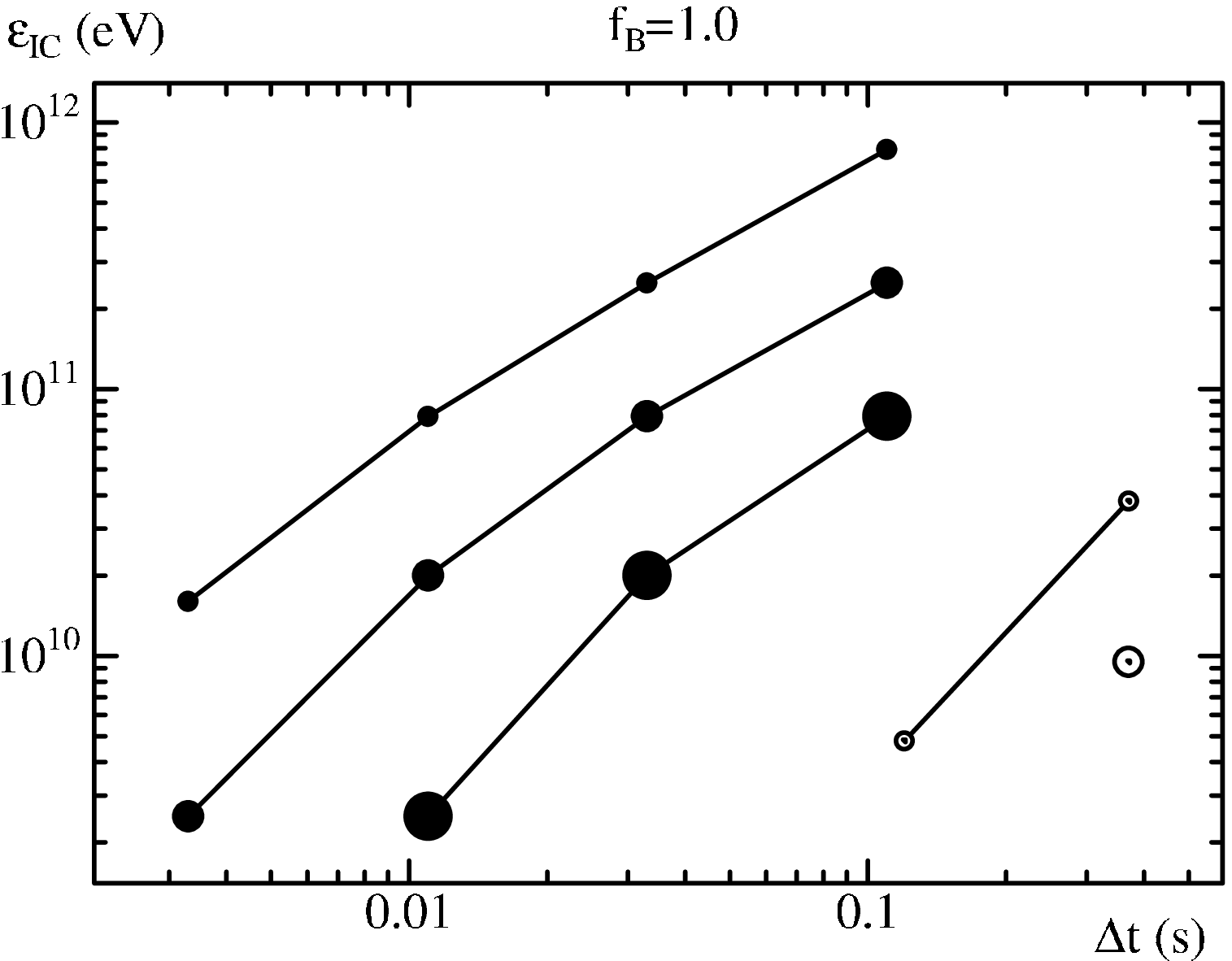}
\plotone{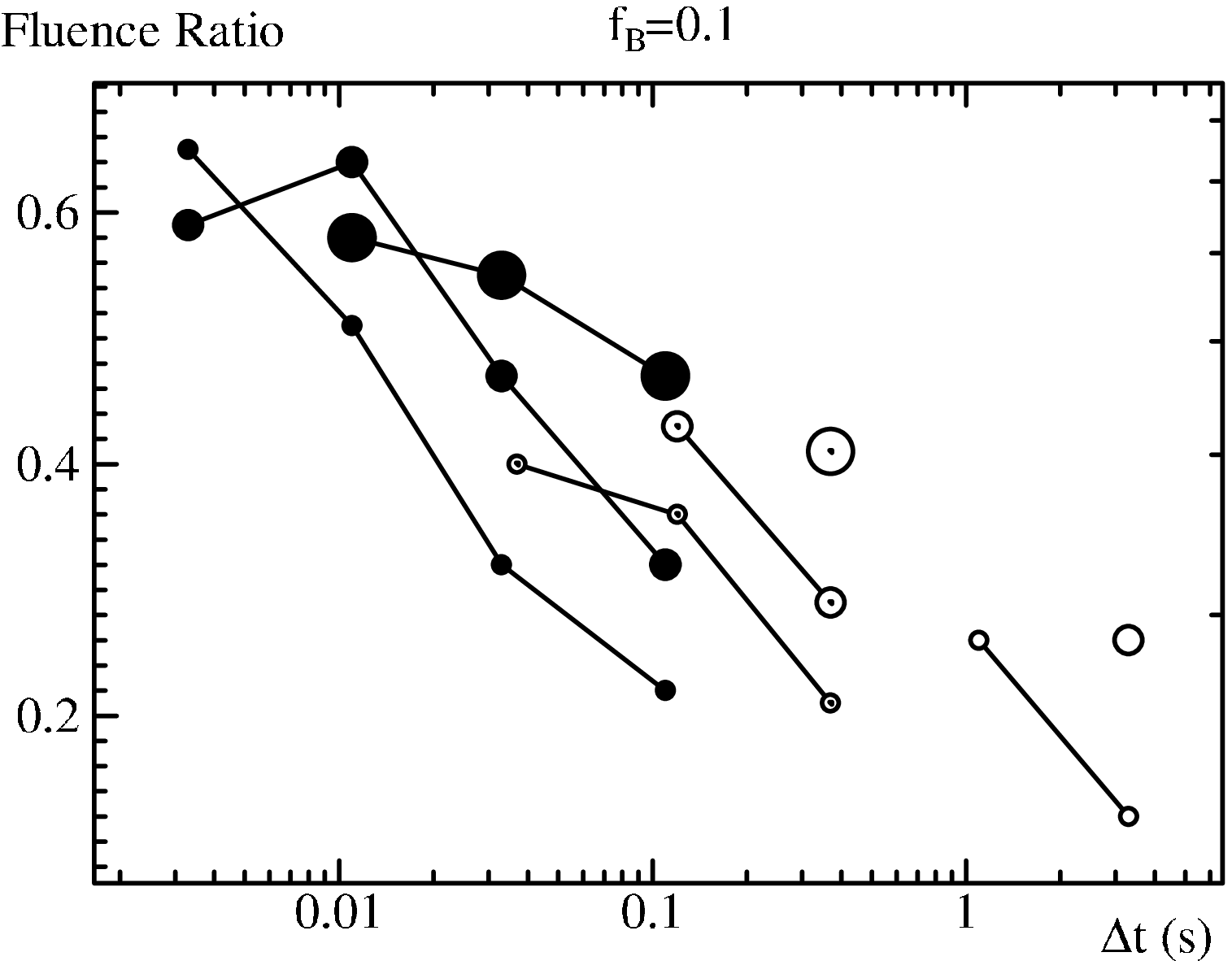}
\plotone{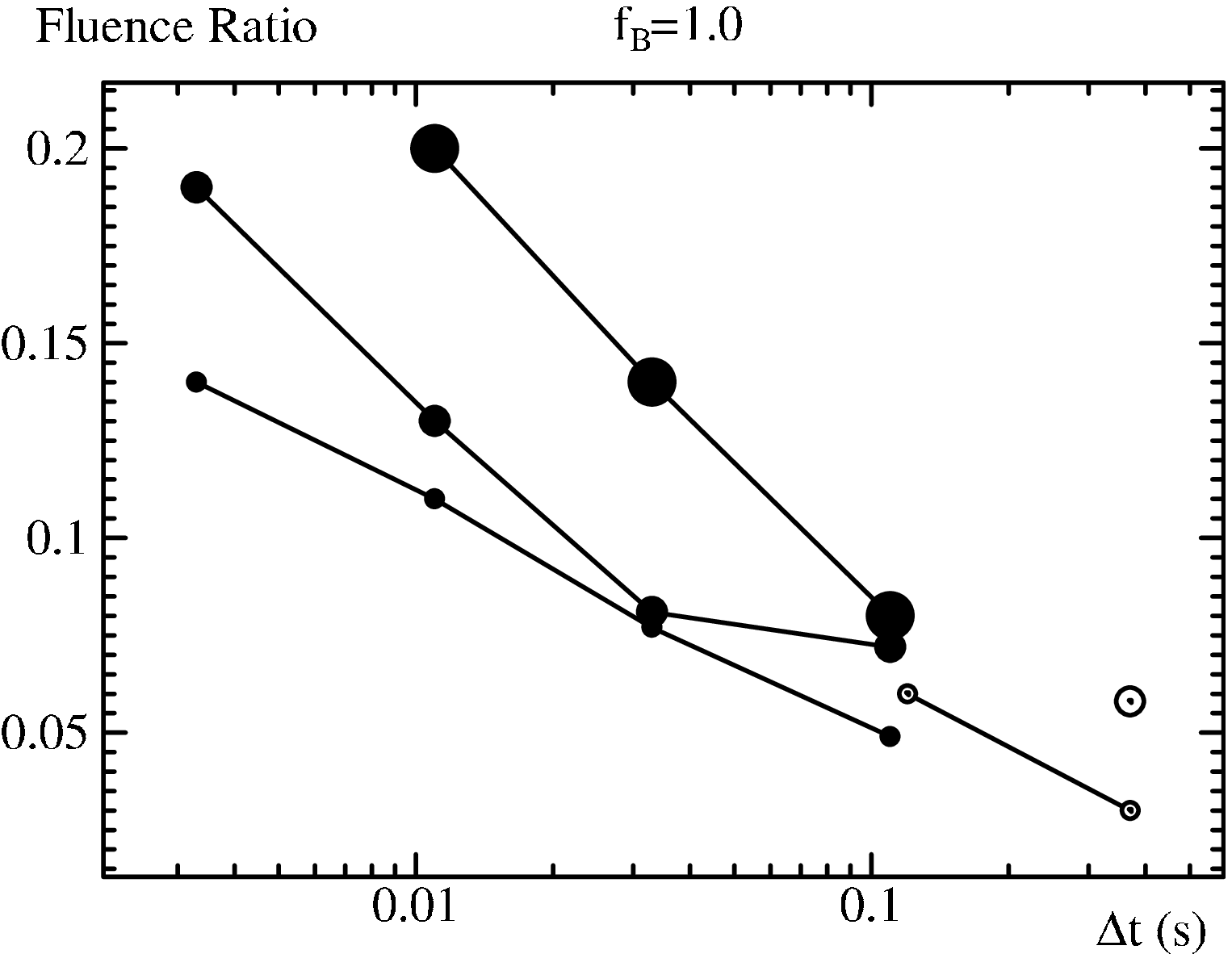}
\caption{
IC peak energy $\varepsilon_{\rm IC}$ (upper panels)
and IC to synchrotron peak fluence ratio (lower panels),
for $f_B=0.1$ (left) and $f_B=1.0$ (right).
Open, dotted and filled circles represent $\Gamma=100$, 300 and 1000,
while small, medium and large circle size correspond to
$E_{\rm sh}=10^{50}$,  $10^{51}$ and $10^{52}$ erg, respectively.
\label{fig:AA}}
\end{figure}

When IC emission is dominant at high-energies (\S \ref{sec:IC}),
the IC peak energy $\varepsilon_{\rm IC}$ and the IC to synchrotron peak fluence
($\varepsilon f(\varepsilon)$) ratio
may allow a useful consistency check of the internal shock model
within the parameter space of the present study (\S \ref{sec:param}).
In Fig. \ref{fig:AA},
we summarize the dependence of these two quantities on $\Delta t$, $E_{\rm sh}$, $\Gamma$ and $f_B$.
Roughly speaking, $\varepsilon_{\rm IC} \propto \Gamma^{4.5} E_{\rm sh}^{-1} \Delta t^{1.5}$,
which is similar to the expression for $\varepsilon_{\rm cut}$ (\S \ref{sec:cutoff}) and sensitive to $\Gamma$.
The peak fluence ratio does not vary monotonically with $E_{\rm sh}$
as it is affected by $\gamma \gamma$ absorption at large $E_{\rm sh}$.
The IC peak is suppressed for larger $f_B$ and disappears for $f_B=30$.


\clearpage

\section{Summary of High-Energy Spectra}
\label{sec:summa}

\begin{figure}[h]
\centering
\epsscale{0.32}
\plotone{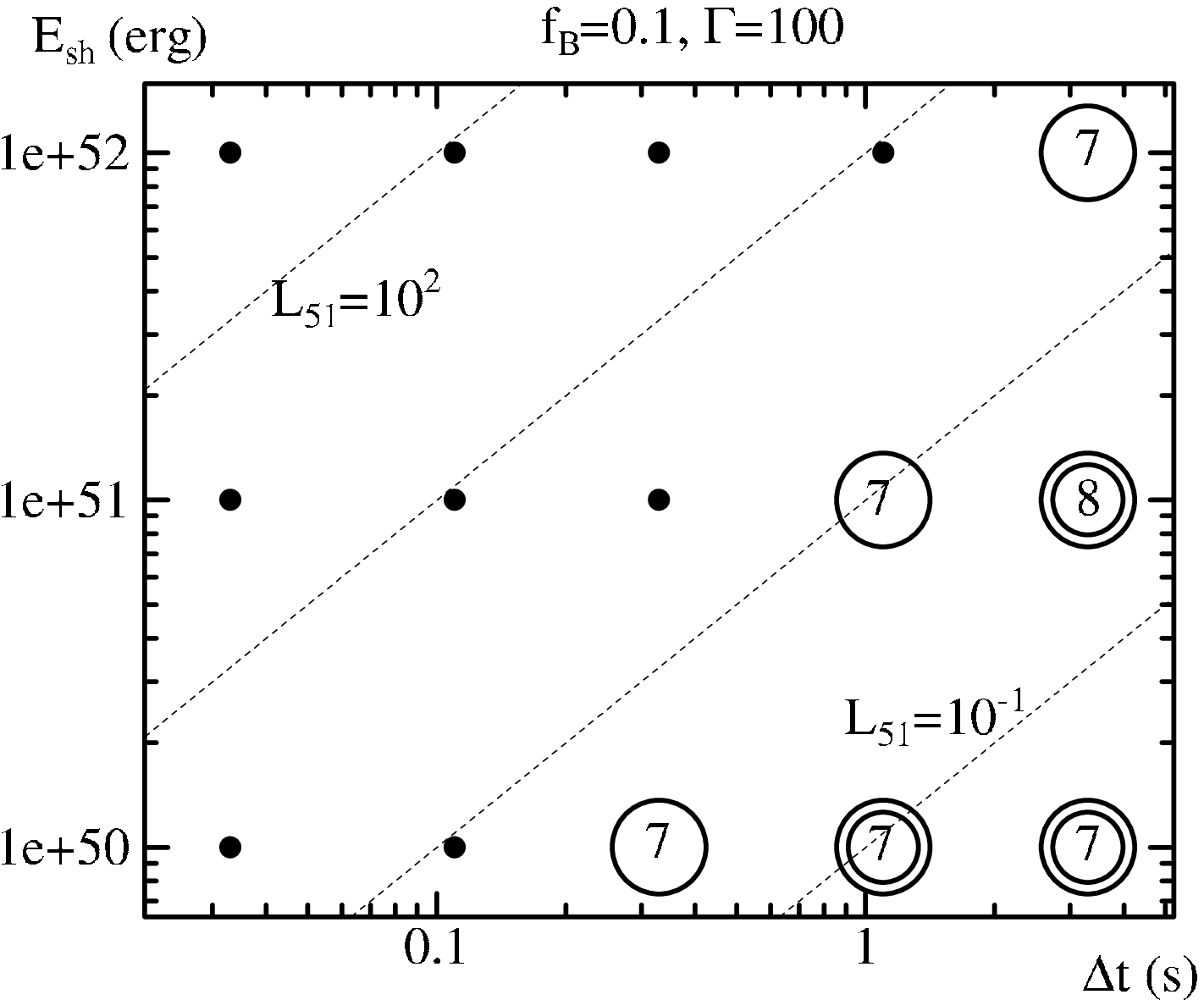}
\plotone{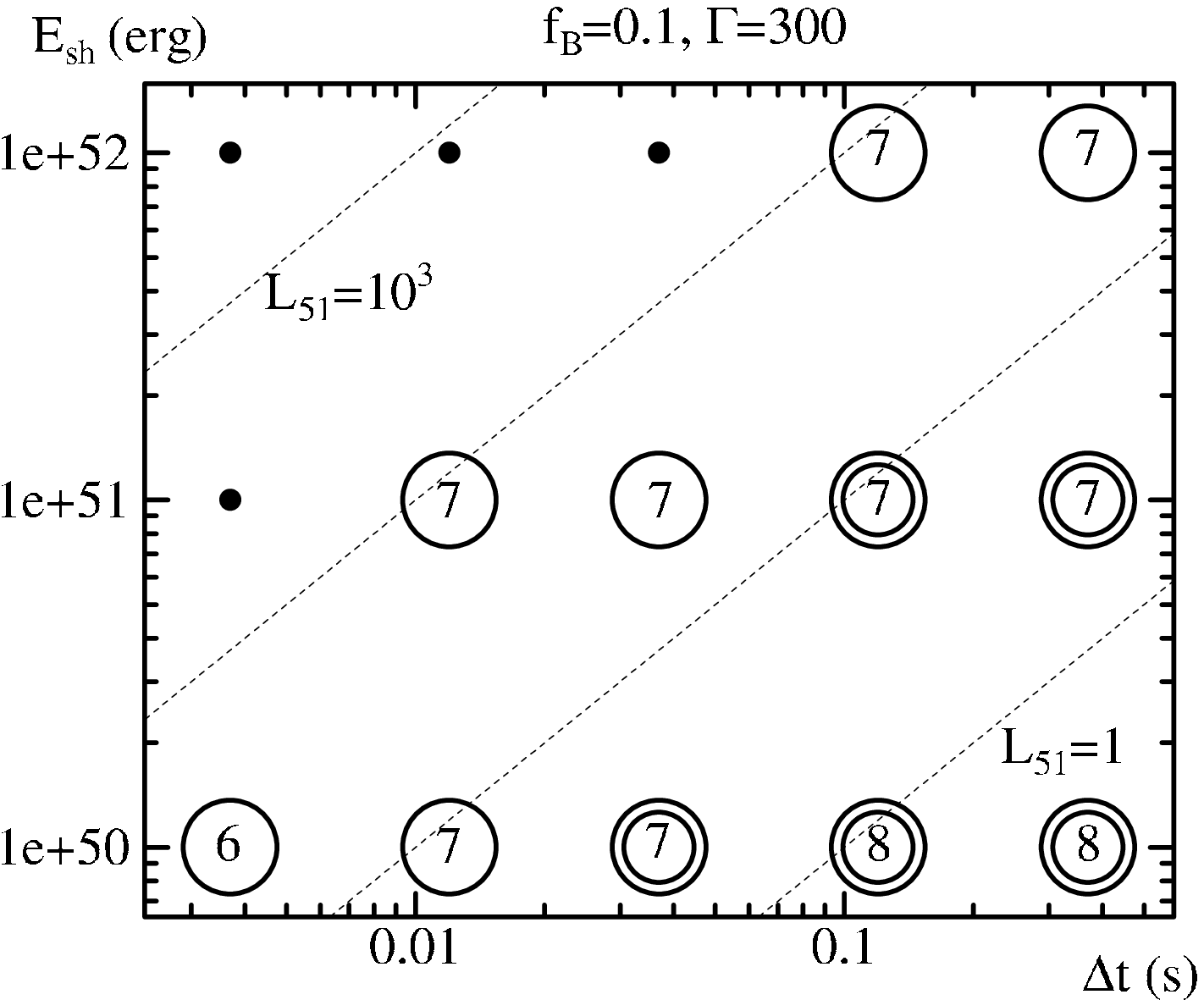}
\plotone{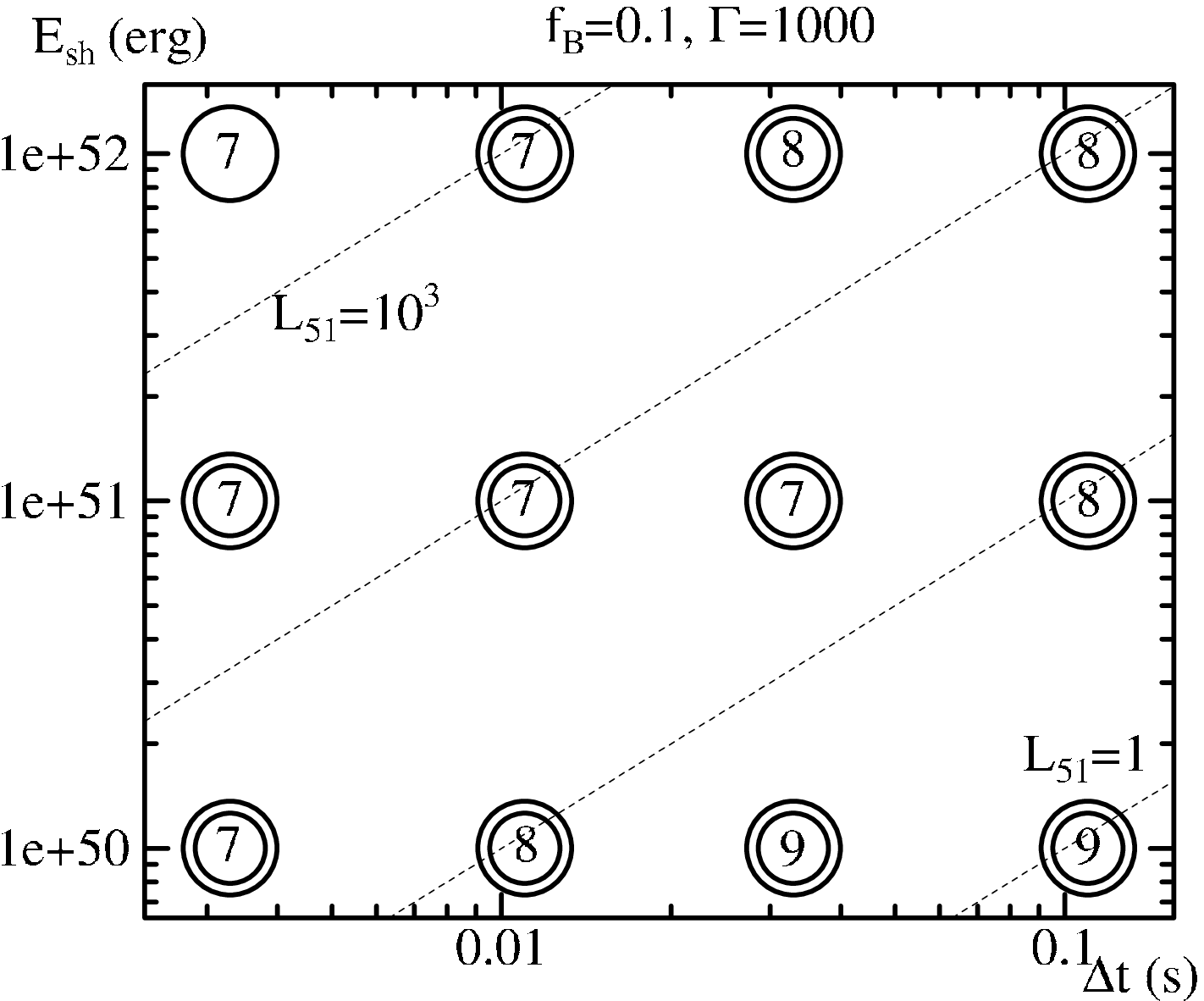}
\plotone{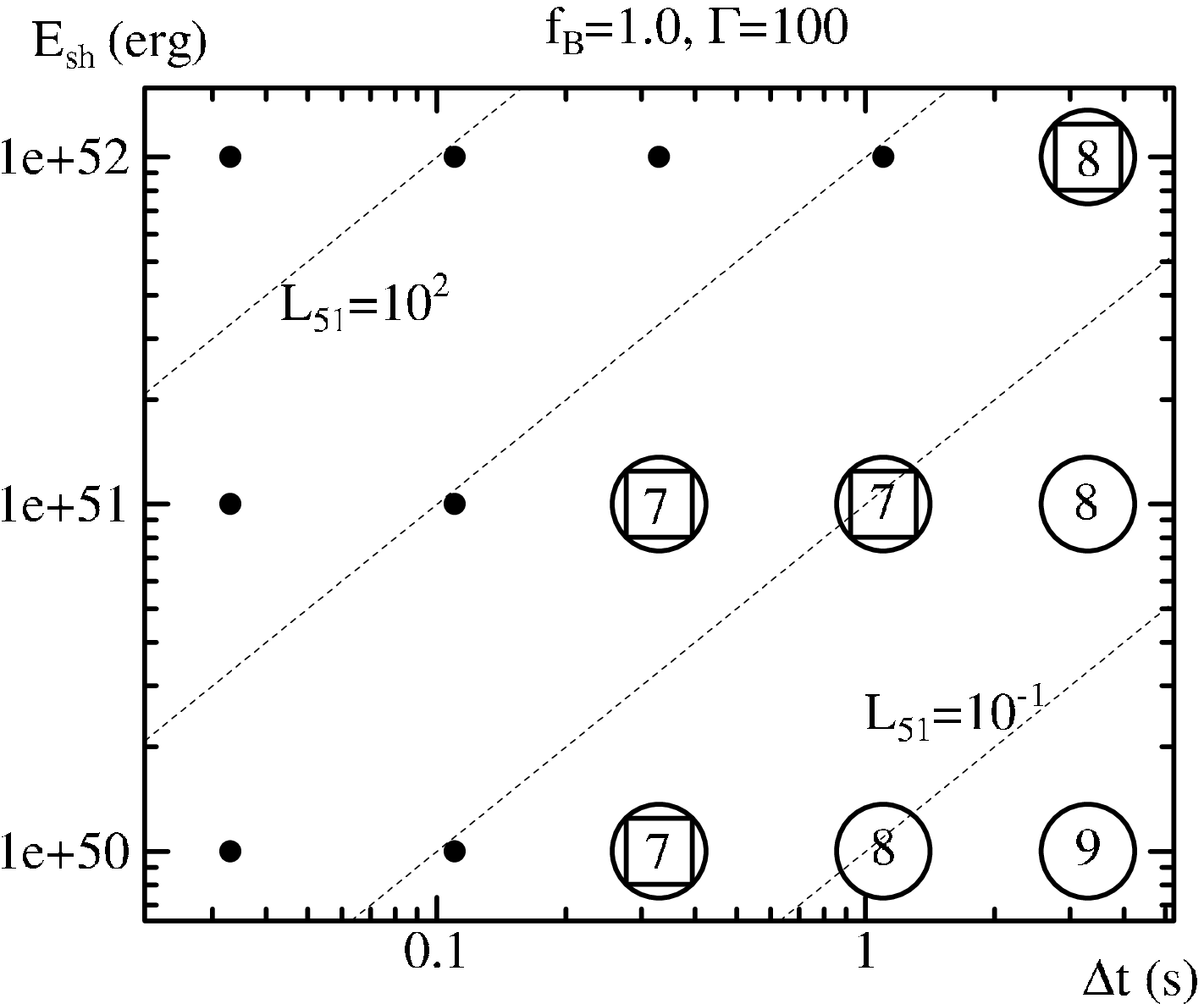}
\plotone{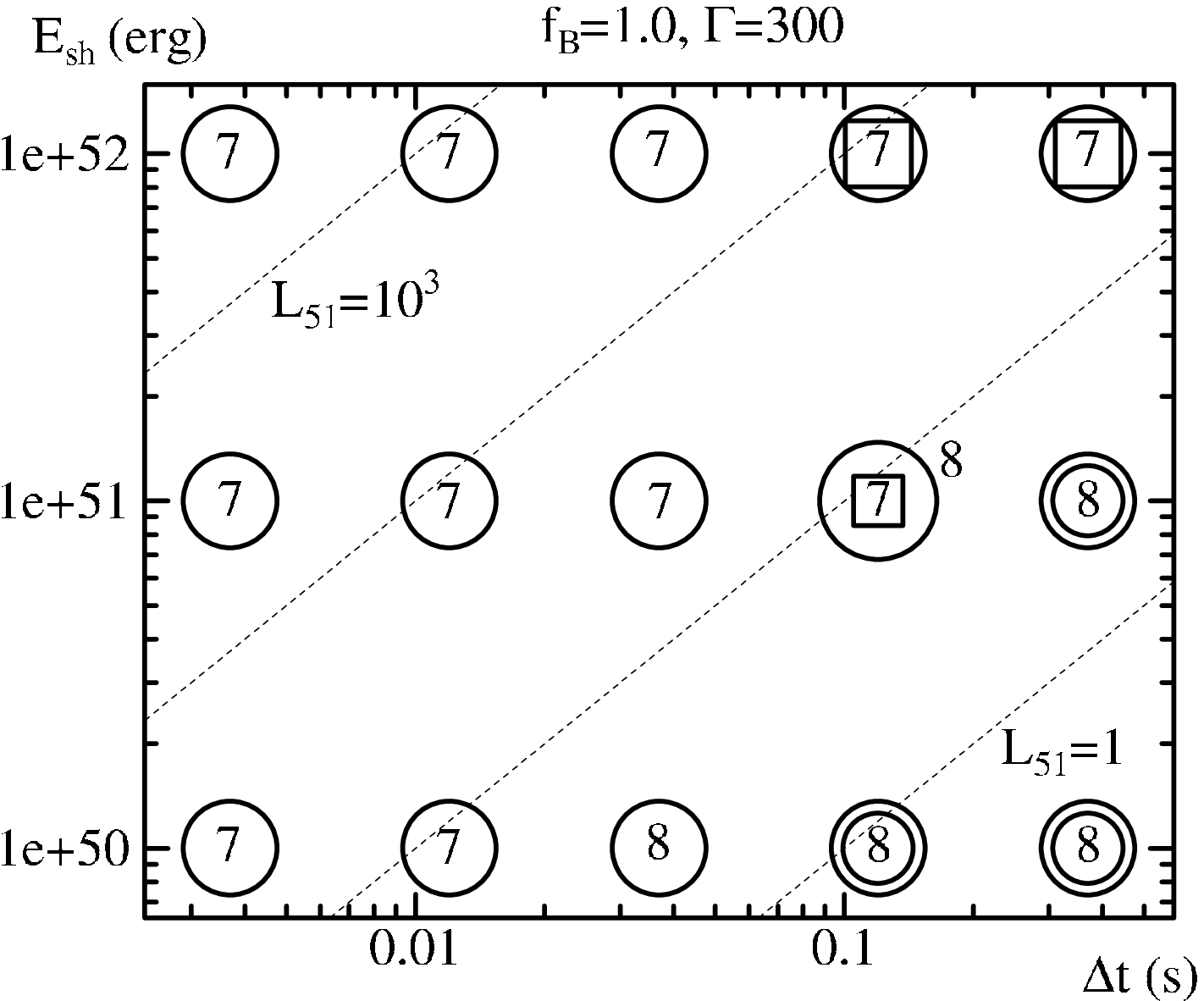}
\plotone{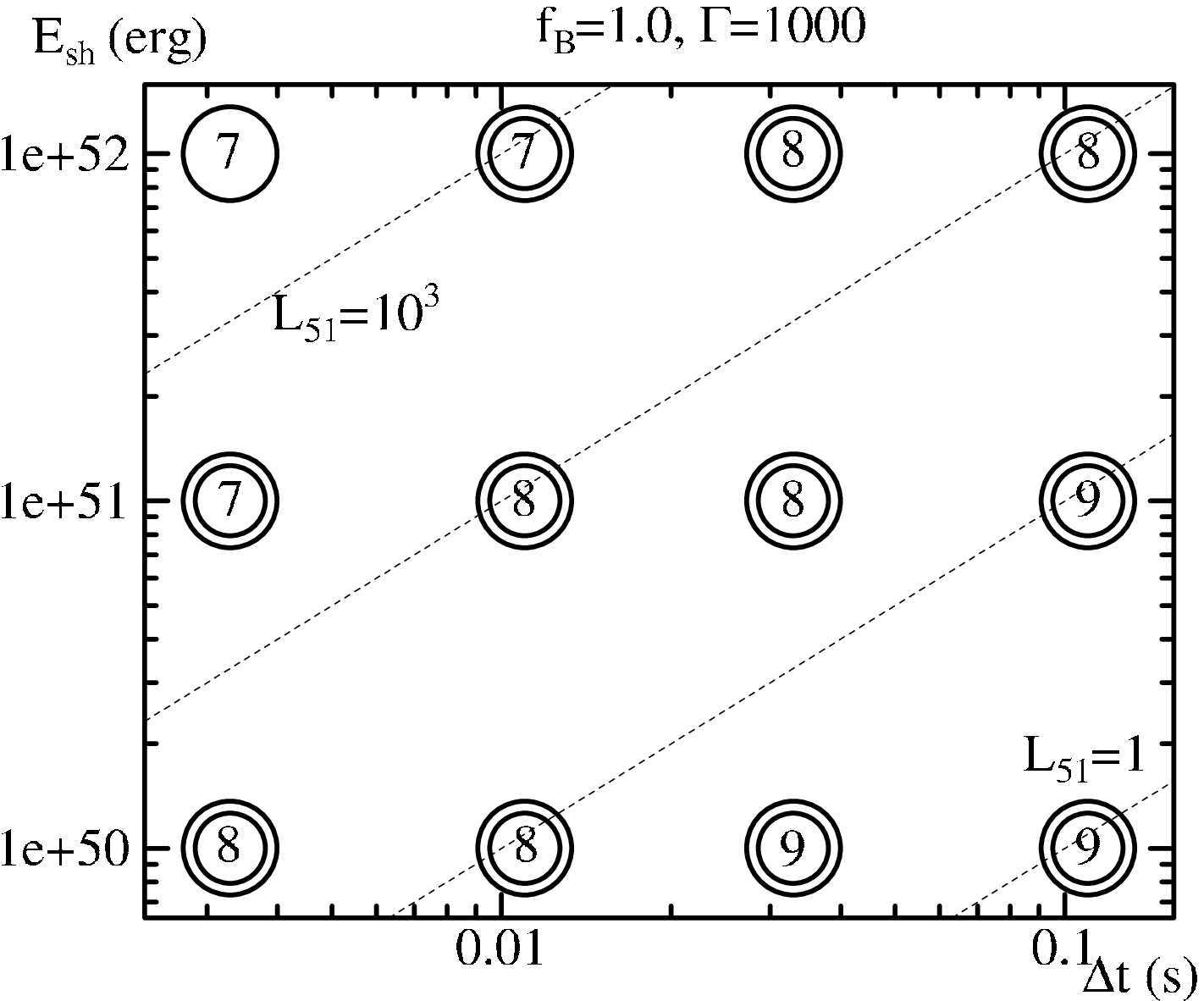}
\plotone{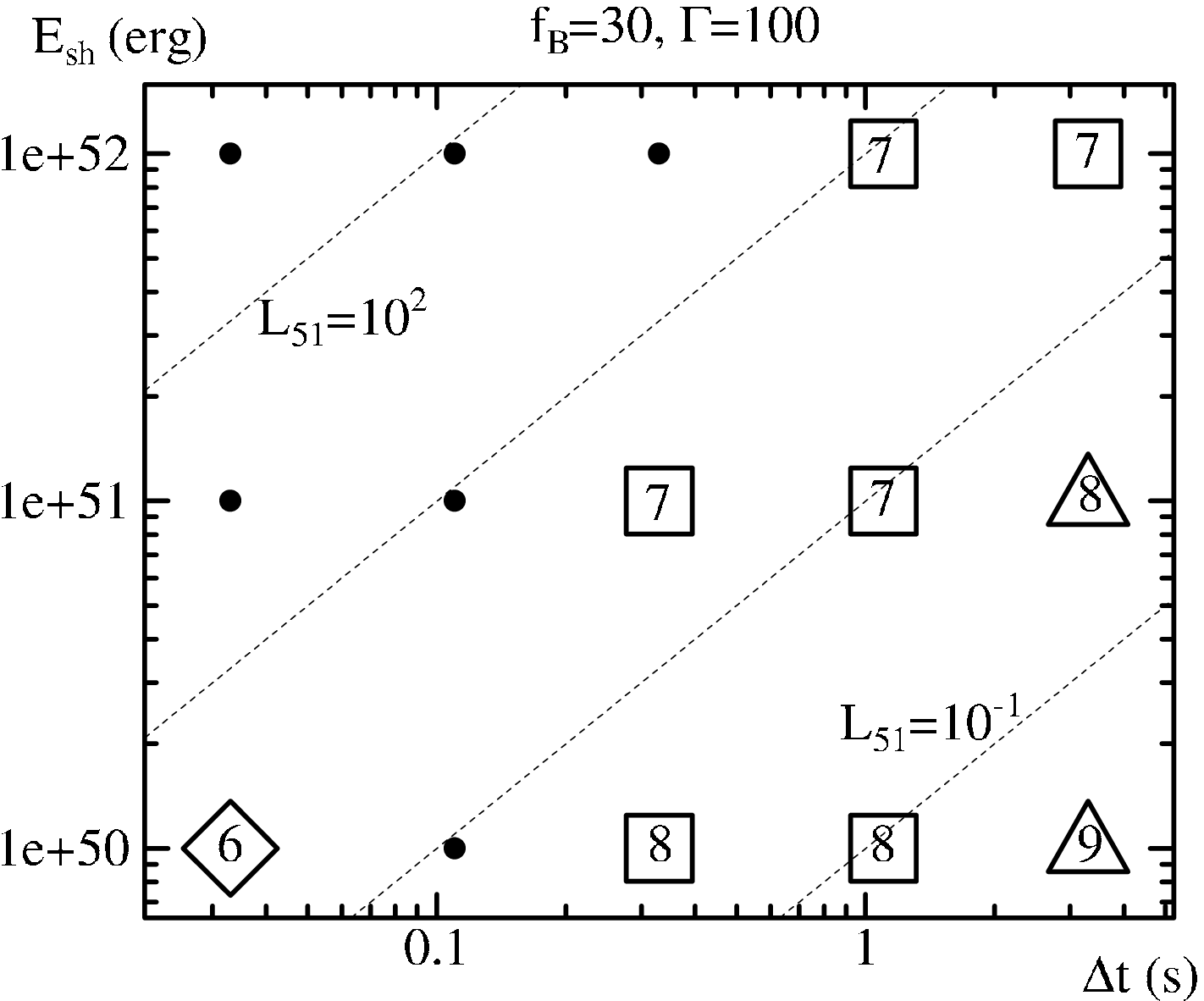}
\plotone{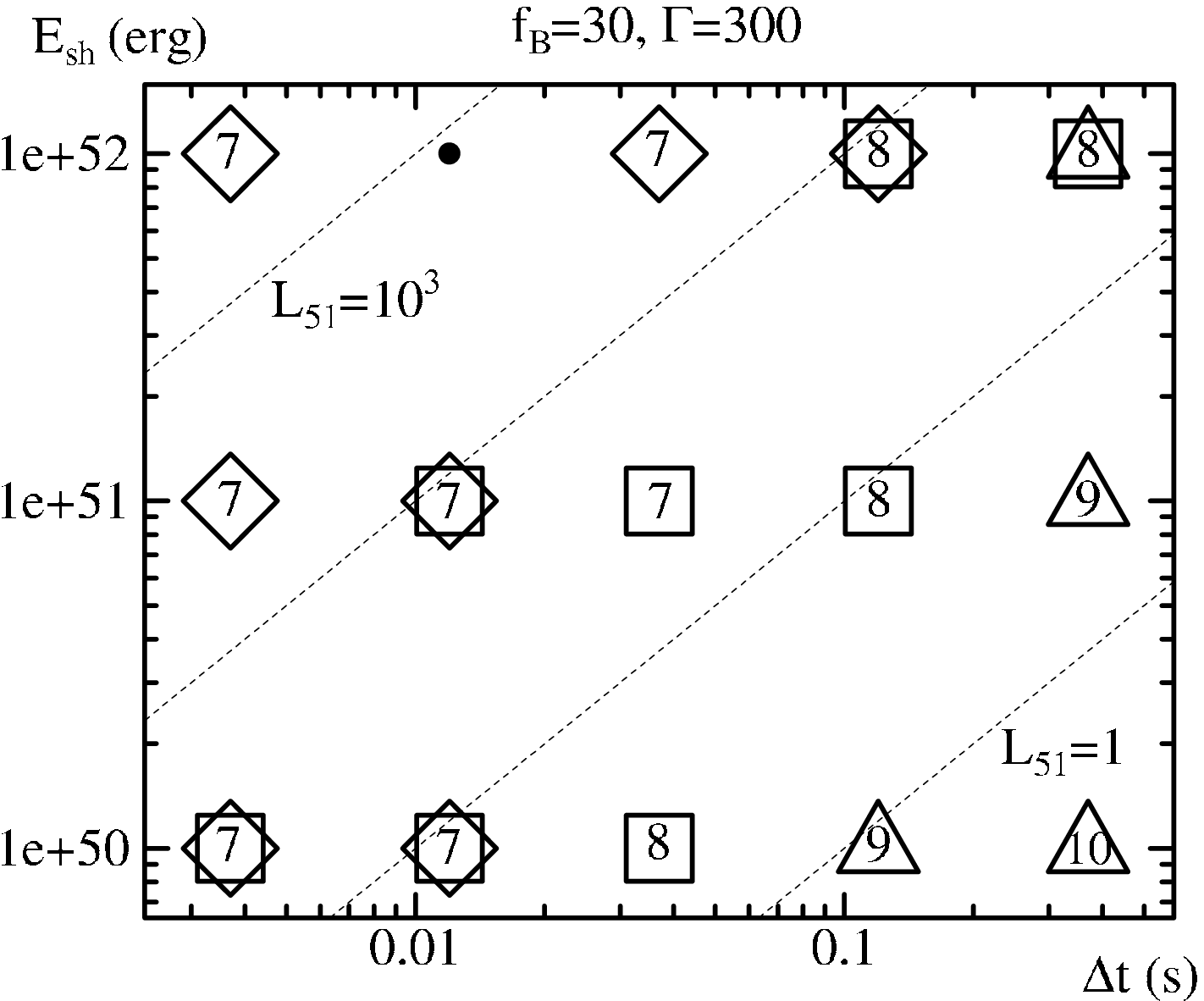}
\plotone{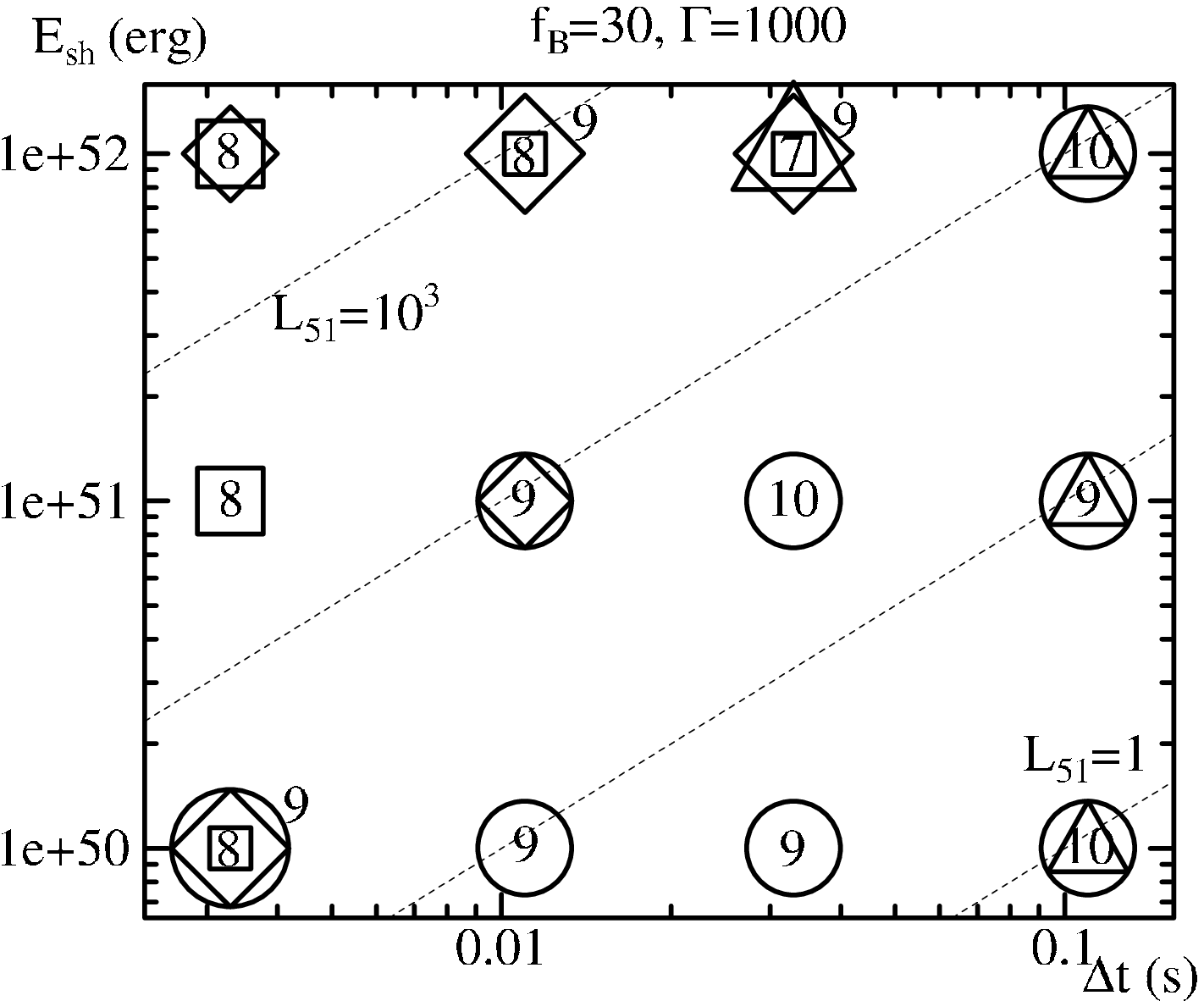}
\caption{
Summary of high-energy spectral components in the $\Delta t$-$E_{\rm sh}$ plane,
with $f_B$ and $\Gamma$ denoted above each panel.
Dotted lines indicate equal luminosity at unit logarithmic intervals
normalized by $L_{51}=L/10^{51} {\rm\ erg\ s^{-1}}$.
The symbols designate the relevant emission process;
circles, rectangles, diamonds and triangles respectively signify
IC, secondary pair synchrotron, muon synchrotron and proton synchrotron emission.
Double circles are cases of IC emission with distinct second peaks.
The number inside the symbol is the photon energy
where the emission component becomes apparent in log eV units.
Two overlapping symbols imply that the respective components occur at  
similar photon energies,
whereas one symbol encircled by another stand for double breaks,
with the corresponding break energies in log eV inscribed beside each  
symbol.
\label{fig:AB}}
\end{figure}

A large variety of high-energy spectra are realized in the current model.
Those with a clear excess above a simple extrapolation of the MeV-band spectrum
are summarized quantitatively in the $\Delta t$-$E_{\rm sh}$ parameter plane in Fig. \ref{fig:AB}.

\section{Neutrino Spectra}
\label{sec:neutrino}

\begin{figure}[h]
\centering
\epsscale{0.5}
\plotone{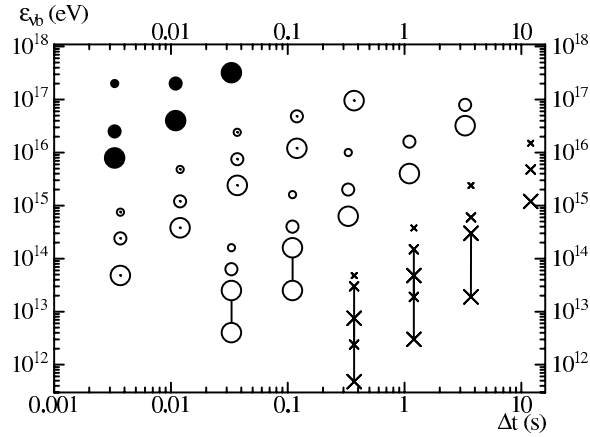}
\caption{
Neutrino break energy $\varepsilon_{\nu b}$ versus $\Delta t$, for $f_B=1.0$.
Crosses, open, dotted and filled circles represent $\Gamma=30$, 100, 300 and 1000,
while small, medium and large symbol size correspond to
$E_{\rm sh}=10^{50}$,  $10^{51}$ and $10^{52}$ erg, respectively.
For cases with two breaks, they are joined by a vertical line.
\label{fig:AC}}
\end{figure}

In the current model, two types of neutrino spectra can occur  
\citep{asa05,asa06}.
When photopion cooling of protons is efficient,
the neutrino number spectrum $\propto \varepsilon_ 
{\nu}^{-2}$ reflecting that of protons,
with a high-energy break at which the synchrotron cooling time of the  
parent pion or muon equals their lifetime \citep{rac98},
and a low energy break at which the photopion cooling time of the  
parent proton equals $t_{\rm exp}$.
In contrast, for inefficient photopion cooling, the spectrum has only  
one break corresponding to the latter
and no portion $\propto \varepsilon_{\nu}^{-2}$.
Fig. \ref{fig:AC} summarizes these characteristic break energies  
$\varepsilon_{\nu b}$
when the neutrino fluence $>10^{-5} {\rm erg\ cm^{-2}}$.
An analytical estimate for the high-energy break gives
$\varepsilon_{\nu b} \propto \Gamma B^ 
{-1} \propto \Gamma R^{1.5} E_{\rm sh}^{-0.5}
\propto \Gamma^4 \Delta t^{1.5} E_{\rm sh}^ 
{-0.5}$,
which agrees with our numerical results.

\clearpage

\clearpage

\end{document}